\documentclass[12pt]{article}
\usepackage{amsmath,amssymb,epsf}

\makeatletter \@addtoreset{equation}{section}
\makeatother

\def\ben{\begin{equation}}
\def\een{\end{equation}}

  \let\n=\nu

\let\C=\Chi

\def\nn{\nonumber} \def\bd{\begin{document}} \def\ed{\end{document}}
\def\ds{\documentstyle} \let\fr=\frac \let\bl=\bigl \let\br=\bigr
\let\Br=\Bigr \let\Bl=\Bigl
\let\bm=\bibitem
\let\na=\nabla
\let\pa=\partial \let\ov=\overline
\newcommand{\be}{\begin{equation}}
\newcommand{\ee}{\end{equation}}
\def\ba{\begin{array}}
\def\ea{\end{array}}
\def\ft#1#2{{\textstyle{\frac{\scriptstyle #1}{\scriptstyle #2}}}}
\def\fft#1#2{\frac{#1}{#2}}
\def\del{\partial}
\def\vp{\varphi}
\def\sst#1{{\scriptscriptstyle #1}}
\def\oneone{\rlap 1\mkern4mu{\rm l}}
\def\td{\tilde}
\def\wtd{\widetilde}
\def\ie{\rm i.e.\ }
\def\dalemb#1#2{{\vbox{\hrule height .#2pt
        \hbox{\vrule width.#2pt height#1pt \kern#1pt
                \vrule width.#2pt}
        \hrule height.#2pt}}}
\def\square{\mathord{\dalemb{6.8}{7}\hbox{\hskip1pt}}}
\newcommand{\ho}[1]{$\, ^{#1}$}
\newcommand{\hoch}[1]{$\, ^{#1}$}
\newcommand{\bea}{\begin{eqnarray}}
\newcommand{\eea}{\end{eqnarray}}
\newcommand{\ra}{\rightarrow}
\newcommand{\lra}{\longrightarrow}
\newcommand{\Lra}{\Leftrightarrow}
\newcommand{\ap}{\alpha^\prime}
\newcommand{\bp}{\tilde \beta^\prime}
\newcommand{\tr}{{\rm tr} }
\newcommand{\Tr}{{\rm Tr} }
\def\0{{\sst{(0)}}}
\def\1{{\sst{(1)}}}
\def\2{{\sst{(2)}}}
\def\3{{\sst{(3)}}}
\def\4{{\sst{(4)}}}
\def\5{{\sst{(5)}}}
\def\6{{\sst{(6)}}}
\def\7{{\sst{(7)}}}
\def\8{{\sst{(8)}}}
\def\n{{\sst{(n)}}}
\def\cA{{{\cal A}}}
\def\cB{{{\cal B}}}
\def\cF{{{\cal F}}}
\def\tV{\widetilde V}
\def\tW{\widetilde W}
\def\tH{\widetilde H}
\def\tE{\widetilde E}
\def\tF{\widetilde F}
\def\tA{\widetilde A}
\def\im{{{\rm i}}}
\def\tY{{{\wtd Y}}}
\def\ep{{\epsilon}}
\def\vep{{\varepsilon}}
\def\R{\rlap{\rm I}\mkern3mu{\rm R}}
\def\bD{{{\bar D}}}

\def\R{\rlap{\rm I}\mkern3mu{\rm R}}
\def\bD{{{\bar D}}}
\def\R{{{\mathbb R}}}
\def\C{{{\mathbb C}}}
\def\H{{{\mathbb H}}}
\def\CP{{{\mathbb C}{\mathbb P}}}
\def\RP{{{\mathbb R}{\mathbb P}}}
\def\Z{{{\mathbb Z}}}
\def\bA{{{\mathbb A}}}
\def\bB{{{\mathbb B}}}
\def\bC{{{\mathbb C}}}
\def\bD{{{\mathbb D}}}
\def\bE{{{\mathbb E}}}
\def\bZ{{{\mathbb Z}}}
\def\Re{{{\frak{Re}}}}
\def\Im{{{\frak{Im}}}}
\def\cosec{{\,\hbox{cosec}\,}}
\def\Gm{{\Gamma_{\!\! -}}}
\def\Gp{{\Gamma_{\!\! +}}}
\def\stan{{standard }}
\def\nonstan{{supernumerary }}

\newcommand{\auth}{
H. L\"u\hoch{\dagger}, Justin F. V\'azquez-Poritz\hoch{\ddagger}
and John E. Wang\hoch{\ast}}

\begin{document}
\begin{flushright}

MIFP-04-11\ \ \ \ \ UK-04-09\ \ \ \ \
UCTP-109-04\ \ \ \ \ USTC-ICTS-04-12\\
{\bf hep-th/0406028}\\
June\  2004
\end{flushright}


\begin{center}

{\large {\bf De Sitter Bounces}}

\vspace{20pt}
\auth

\vspace{20pt} {\hoch{\dagger}\it George P. and Cynthia W. Mitchell
Institute for Fundamental Physics,\\ Texas A\& M University,
College Station, TX 77843-4242, USA}

\vspace{10pt} {\hoch{\dagger}\it Interdisciplinary Center for
Theoretical Study, \\ University of Science \& Technology of China,
Hefei, Anhui 230026, China}

\vspace{10pt} {\hoch{\ddagger}\it Department of Physics and Astronomy,\\
University of Kentucky, Lexington, KY 40506}

\vspace{10pt} {\hoch{\ddagger}\it Department of Physics,\\
University of Cincinnati, Cincinnati OH 45221-0011}

\vspace{10pt} {\hoch{\ast}\it Department of Physics,\\
Harvard University, Cambridge MA 02138}

\vspace{10pt} {\hoch{\ast}\it Department of Physics,\\
National Taiwan University, Taipei 106, Taiwan}

\vspace{30pt}

\underline{ABSTRACT}
\end{center}

By analytically continuing recently-found instantons we construct
time-dependent solutions of Einstein-Maxwell de Sitter gravity
which smoothly bounce between two de Sitter phases.  These
deformations of de Sitter space undergo several stages in their
time evolution. Four and five-dimensional de Sitter bounces can be
lifted to non-singular time-dependent solutions of M-theory.

\pagebreak

\setcounter{page}{1}


\section{Introduction}

Cosmological bounces may provide an alternative to inflation, in
terms of addressing the homogeneity problem and providing a causal
mechanism of structure formation. Instead of an inflationary phase
of quasi-exponential expansion, a bounce could have a long period
of slow contraction, during which currently observed cosmological
scales would be well inside the Hubble radius. Then, for example,
there would be ample time for the observed homogeneity to occur
through causal microphysics.

String-based models, such as Pre-Big-Bang cosmology
\cite{veneziano1,veneziano2} and the Ekpyrotic scenario
\cite{khoury}, have led to an increased interest in cosmological
bounces. In fact, it was long believed that stringy corrections
and quantum loop effects were required to smooth a cosmological
singularity out to a smooth bounce \cite{antoniadis}.  This was
reinforced when cosmological solutions were later found which
turned out to all be singular \cite{ovrut1,lu1,ovrut2,lu2}; these
solutions are now generally classified as S(pacelike)-branes
\cite{gutperle}.

The construction of classical non-singular time-dependent
solutions is therefore of interest. One approach to resolve a
cosmological singularity is for the cosmological flow to connect
early and late-time de Sitter spacetimes \cite{cvetic}. It has
been found in \cite{adsbh,desitter} that AdS black holes can be
analytically continued to non-singular cosmological solutions. In
four dimensions, the resulting time-dependent solution smoothly
interpolates between dS$_2\times S^2$ and a dS$_4$-type geometry
with a boundary of $S^2\times S^1$.  These solutions are not
supersymmetric, although a subset solves the first-order equations
that coincide with the BPS equations for a non-compact
$R$-symmetry. Also, these solutions are in a minimum of the
potential with no tachyonic directions. This indicates that these
smooth cosmological solutions may be stable \cite{cvetic}. The
more general class of these solutions arise from the second-order
equations of motion and can be fine-tuned to describe an expanding
universe whose expansion rate is significantly larger in the past
than in the future, providing a cosmological model with no
singularities.  These four-dimensional solutions have been lifted
to non-singular S-brane configurations in eleven dimensions
\cite{lvpds,lvpsmooth}.

More recently, non-singular S-branes have also been shown to arise
from the analytical continuation of diholes \cite{strominger},
Kerr black holes \cite{wang,quevedo} and rotating $p$-branes
\cite{lvp}. Generically these solutions either smoothly
interpolate from a warped product of two-dimensional de Sitter
spacetime and an internal space to Minkowski spacetime, or
smoothly bounce between two phases of Minkowski spacetime. These
solutions explicitly demonstrate that certain cosmological
singularities can be smoothed out to yield smooth bounces without
the inclusion of stringy or quantum effects. Also unlike the
resolution of p-brane singularities, which tend to require
additional flux fields \cite{strassler}, these cosmological
bounces rely only on classical gravitational physics and not
matter fields.

In constructing a cosmological bounce model, it is important to
take into account the late-time behavior. Astrophysical evidence
suggests that our universe is currently in a de Sitter phase
\cite{astro1,astro2}. In this paper, we construct de Sitter
bounces by analytically continuing the de Sitter instantons found
in \cite{lpp}. In certain cases, we find that the four-dimensional
de Sitter bounce satisfies BPS constraints, which implies that the
solution might be stable. Interestingly, the bounce can also be
fine-tuned to reduce the effect of the cosmological constant for
an intermediate period of time, which might result in a longer
period of slow contraction. In this case, the cosmological bounce
could be an alternative to inflation.

The rest of this paper is organized as follows. In the next
section, we analytically continue de Sitter instantons to
time-dependent solutions which smoothly bounce between two de
Sitter phases. In section 3, we generalize the four-dimensional de
Sitter bounce to include electric and magnetic charges. This
solution can be analytically continued from an AdS
Reissner-N\"{o}rdstrom Taub-NUT solution. In section 4, we lift
the four and five-dimensional de Sitter bounces to
eleven-dimensional supergravity and type IIB theory, respectively.
Conclusions are presented in section 5.

\section{From instantons to bounces}

We begin with the recently-discussed instanton solutions to
Euclidean de Sitter gravity, found in \cite{lpp}
\footnote{Low-lying examples were also obtained in \cite{mann}}
\be
ds_d^2=\fft{(1-r^2)^n}{P(r)}\,dr^2+\fft{c^2\,P(r)}{(1-r^2)^n}\,
(d\psi-2A)^2+c\,(1-r^2)\,d\Sigma_{2n}^2+
\fft{m-1}{\Lambda-\lambda\,c^{-1}}\,
r^2\,d\Omega_m^2, \label{metricmn} \ee
where $c$ is an arbitrary integration constant, $\lambda$ is the
cosmological constant of the $2n$-dimensional Einstein-K\"{a}hler
space $\Sigma_{2n}$ with the metric $d\Sigma_{2n}^2$ and $\Lambda$
is the cosmological constant of the $d=2n+m+2$-dimensional metric
$ds_d^2$. After rescaling, these constants only appear in the
dimensionless combination
\be \nu\equiv \fft{c\,\Lambda}{\lambda}\,. \ee
In our notation, $d\Omega_m^2$ is the metric of a unit sphere
$S^m$ and $A$ is the potential for the K\"{a}hler form on
$d\Sigma_{2n}^2$, such that the K\"{a}hler two-form is given as
$J=dA$. The function $P(r)$ is given by
\be P(r)=\fft{\Lambda}{m-1}\,_2F_1
(-n-1,\ft{m-1}{2};\ft{m+1}{2};r^2)- \fft{\lambda\,c^{-1}}{m-1}
\,_2F_1 (-n,\ft{m-1}{2};\ft{m+1}{2};r^2)+\mu\,r^{1-m}\,, \ee
where $_2F_1$ denotes the standard hypergeometric function.  The
parameter $\mu$ is an integration constant. Additional details of
this solution can be found in \cite{lpp}.  For easy reference, we
write the first few hypergeometric functions explicitly as
polynomials in $r$:
\begin{eqnarray}
_2F_1 (-1,\ft{m-1}{2};\ft{m+1}{2};r^2) &=&\frac{1}{m-1}
-\frac{r^2}{m+1}\\
_2F_1 (-2,\ft{m-1}{2};\ft{m+1}{2};r^2) &=&\frac{1}{m-1}
-\frac{2r^2}{m+1}+\frac{r^4}{m+3}\\
_2F_1 (-3,\ft{m-1}{2};\ft{m+1}{2};r^2) &=&\frac{1}{m-1}
-\frac{r^2}{m+1}+\frac{3r^4}{m+3}-\frac{r^6}{m+5} \ .
\end{eqnarray}

Consider the analytical continuation of the coordinates and
parameters given by
\be r\rightarrow {\rm i}\,t\,,\qquad \mu \rightarrow {\rm
i}^{m-1}\,\mu\,,\qquad d\Omega_m^2\rightarrow -dH_m^2\,, \ee
The metric (\ref{metricmn}) becomes
\be ds_d^2=-\fft{(1+t^2)^n}{\td P(t)}\,dt^2+\fft{c^2\,\td
P(t)}{(1+t^2)^n}\, (d\psi-2
A)^2+c\,(1+t^2)\,d\Sigma_{2n}^2+\fft{(m-1)\nu}{\Lambda\,(\nu-1)}\,
t^2\,dH_m^2\,, \label{bouncemn} \ee
where $dH_m^2$ is the metric of a unit hyperbolic $m$-plane and
$\td P(t)$ is given by
\be \td P(t)=\fft{\Lambda}{m-1}\,_2F_1
(-n-1,\ft{m-1}{2};\ft{m+1}{2};-t^2)- \fft{\lambda\,c^{-1}}{m-1}
\,_2F_1 (-n,\ft{m-1}{2};\ft{m+1}{2};-t^2)+\mu\,t^{1-m}\,.
\label{Pmn} \ee

     Asymptotically as $t\rightarrow \pm\infty$, the function $\td P(t)$
takes the simple form
\be \td P(t)=\fft{\Lambda}{2n+m+1}(t^2 + 1)^{n+1}\, \Bigl[1 +
{\cal O}(\fft{1}{t^2})\Bigr]\,. \ee
For the case where $d\Sigma_{2n}^2$ is the metric of $\CP^n$, the
spacetime is asymptotically de Sitter with the boundary
$S^{2n+1}\times H^m$, provided that $\nu>1$.  For the case of
$\nu=1$, $H^m$ is blown up to a torus $T^m$ and when $\nu<1$,
$H^m$ has to be analytically continued back to a sphere $S^m$.

        For the short-distance behavior, we consider three different
cases: $m\ge 2$, $m=1$ and $m=0$.  For $m\ge 2$, the absence of a
curvature singularity at $t=0$ requires that $\mu=0$.  In
addition, if $\nu>1$ then the function $\td P(t)$ is
positive-definite and the metric describes a bounce between two
asymptotic de Sitter spacetimes.  For $\nu=1$, $P\ge 0$ and this
function has a second-order zero at $t=0$, near which the metric
becomes
\be ds_d^2 = -\fft{d\tau^2}{\ft12P''(0)} + e^{2\tau}\, \Bigl(
\ft12 c^2\, P''(0)\, d\psi^2 + dx^i\, dx^i \Bigr) +
\fft{\lambda}{\Lambda}\, d\Sigma_{2n}^2\,, \ee
for $\tau=\log t\rightarrow -\infty$.  In this limit the metric
can be viewed as the direct product dS$_{m+2}\times \Sigma_{2n}$.
The spacetime interpolates between dS$_{m+2}\times \Sigma_{2n}$ at
$t=0$ to dS$_{2n+m+2}$ in the infinite future. There are no closed
timelike curves for $\nu\ge 1$.

When $\nu<1$, $dH_m^2$ must be analytically continued back to
$-d\Omega_m^2$.  Furthermore, the function $\td P(t)$ will have
two first-order zeroes at  $t=\pm t_0$.  Between these two zeroes
the function $\td P(t)$ is negative and the periodic coordinate
$\psi$ becomes timelike; when $\nu<1$, there are stationary
regions with closed timelike curves.

\begin{figure}[ht]
   \epsfxsize=4.0in \centerline{\epsffile{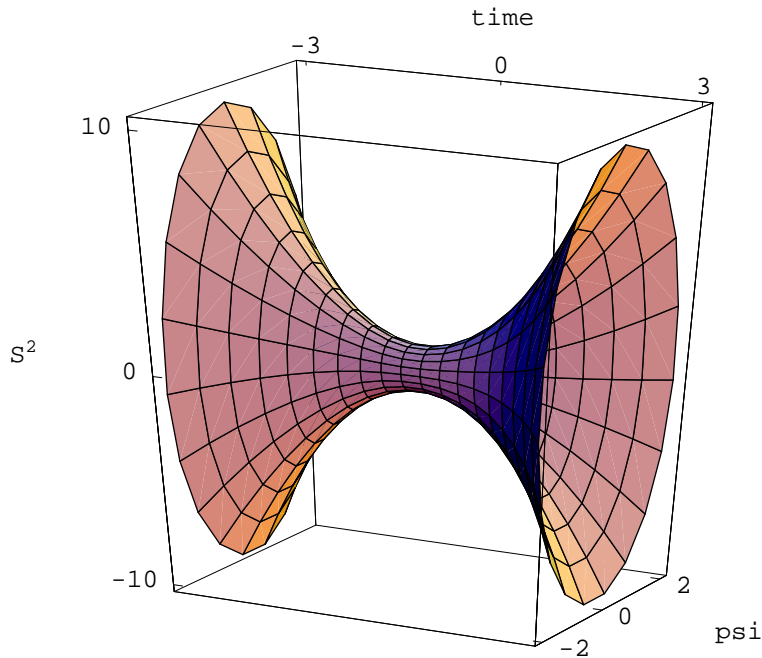}}
   \caption[FIG. \arabic{figure}.]{Bounce between two four-dimensional
   asymptotically de Sitter spacetimes; $\nu=2/3$ and $\mu=0$.}
   \label{bounce1}
\end{figure}

     Next, we turn to the case $m=1$. After scaling the metric, we obtain
\be ds^2 = -\fft{(1+t^2)^n}{\td P(t)}\,dt^2+\fft{c^2\,\td
P(t)}{(1+t^2)^n}\,
(d\psi-2A)^2+c\,(1+t^2)\,d\Sigma_{2n}^2+P_0^{-1}\,t^2\, d\phi^2\,,
\label{bounce1n} \ee
\be \td P(t) =\fft{\Lambda}{2(n+1)}\,[(1+t^2)^{n+1}-1]+\mu\,.
\label{P1} \ee
For $\mu >0$, the function $P$ is positive-definite and hence the
metric describes a bounce between two asymptotic dS$_{2n+3}$
spacetimes at $t\rightarrow \pm \infty$.   When $\mu=0$, the
function $\tilde{P}(t)$ has a second-order zero at $t=0$. In this
case, the metric interpolates between dS$_3\times \Sigma_{2n}$ at
$t=0$ to dS$_{2n+3}$ in the infinite future.  For $\mu<0$,
$\tilde{P}$ can be negative and there are regions with closed
timelike curves.

      Finally, we consider the case of $m=0$. Let us define the
function $Q(t)$, given by $Q(t)=\td P(t,\mu=0)$. The requirement
that $\td P(t)$ is positive-definite implies that $Q(t)$ must also
be positive-definite. The function $Q(t)$ is non-negative provided
that $\nu_0\le \nu\le 1$, where $\nu_0$ is the only positive root
of the polynomial equation $Q(\sqrt{\ft1{\nu_0}-1})=0$.  For
example, for $n=1$, we have $\nu_0=\ft14$; for $n=2$, we have
$\nu_0=(2 + \sqrt{10})/12=0.43019$; for $n=3$, we have
$\nu_0=0.543067$; for $n=4$, we have $\nu_0=0.619223$.  It is
clear that $\nu_0$ approaches 1 as $n$ approaches infinity.
\begin{figure}[ht]
   \epsfxsize=4.0in \centerline{\epsffile{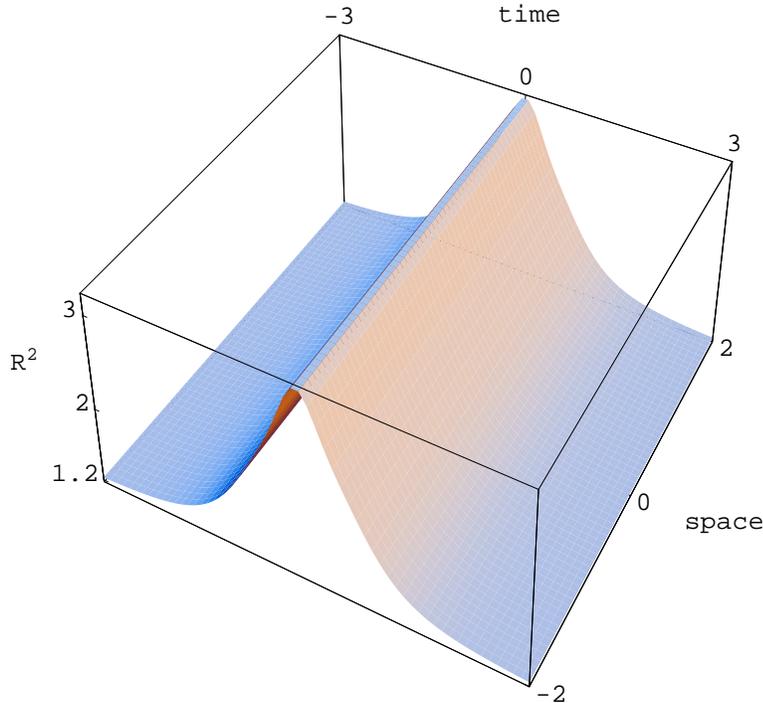}}
   \caption[FIG. \arabic{figure}.]{$R^{mnpq}R_{mnpq}$ corresponding to
   the de Sitter bounce in Figure 1.}
   \label{bounce2}
\end{figure}
        Having determined the range of $\nu$ such that $Q(t)$ is
non-negative, we can ask what is the range of $\mu$ so $\td P(t)$
is also non-negative. We find that we need $|\mu|\le \mu_0$, where
$\mu_0=\Lambda\, Q(t_0)/t_0$ and $t_0=\sqrt{\ft1{\nu}-1}$. Thus,
for $\nu_0<\nu<1$ and $|\mu|<\mu_0$, the function $\td P(t)$ is
positive-definite and the metric describes a bounce between two
asymptotically de Sitter spacetimes. In Figure 1, we plot the
radii of the $S^2$ and fibre bundle direction $\psi$ as a function
of time for the four-dimensional ($n=1$ and $m=0$) de Sitter
bounce. We have taken the parameter values $\nu=2/3$, $\mu=0$ and,
in the Figure, have chosen the convention that time lies along a
horizontal axis. The important property to note is that the radii
decrease towards the middle of the bounce but never completely
vanish. This corresponds to $R^{mnpq}\,R_{mnpq}$ being
well-behaved, as shown in Figure 2. It approaches a finite maximum
value in the middle of the bounce and then asymptotes to a lower
but positive value corresponding to the asymptotic de Sitter
phases.

For $\nu=1$ or $\nu=\nu_0$, with $\mu=0$, the function $\td P(t)$
has a second-order zero at $t_0=1$ or $t_0=\sqrt{\ft1{\nu_0}-1}$,
respectively and so the solution is dS$_2\times \Sigma_{2n}$ at
$t_0$.  The solution flows from dS$_2\times \Sigma_{2n}$ at time
$t_0$ to dS$_{2n+2}$ in the infinite future. If either $\mu$ or
$\nu$ lie outside the range detailed above, then although the
solution is free from curvature singularities, the solution
contains regions with closed timelike curves.  We illustrate this
in Figure 3 for the four-dimensional de Sitter bounce for the case
of $\nu=1/8$ and $\mu=0$.  Again, time runs along the horizontal
axis going from left to right. The axis coming out of the page is
the radius of the $S^2$ as a function of time, which never
vanishes. The radius of the fibred direction $\psi$ is plotted
along the vertical direction and we have oriented the plot to
illustrate that this radius does vanish at four different points,
corresponding to Cauchy horizons separating dynamical and
stationary regions. There is a time-dependent region in the middle
of the bounce, followed and preceded by stationary regions
containing closed timelike curves. These, in turn, are followed
and preceded by time-dependent regions which asymptote to de
Sitter spacetime.
\begin{figure}[ht]
   \epsfxsize=4.0in \centerline{\epsffile{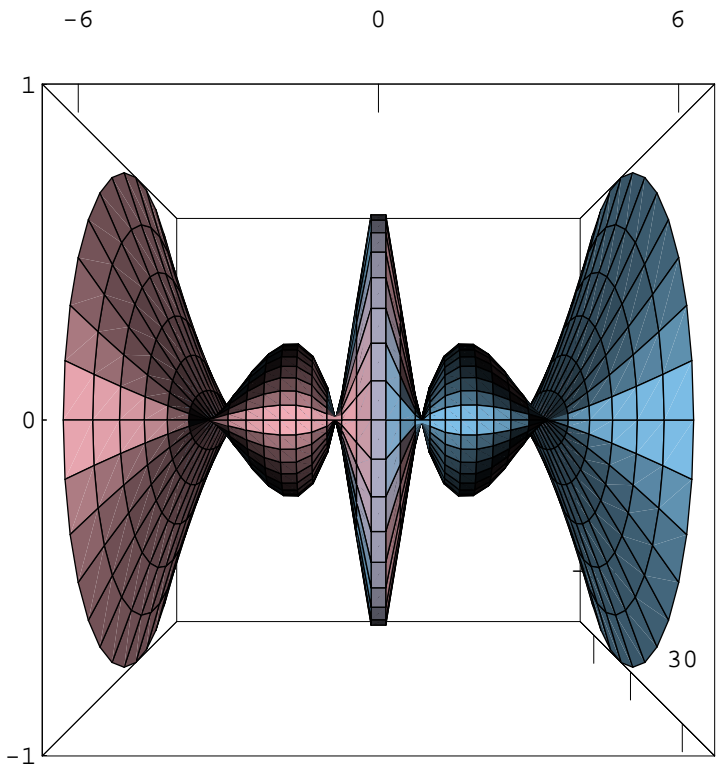}}
   \caption[FIG. \arabic{figure}.]{Bounce between two four-dimensional
   asymptotically de Sitter spacetimes with stationary regions
containing closed timelike curves; $\nu=1/8$ and $\mu=0$.}
   \label{bounce3}
\end{figure}

The time-dependent solutions discussed in this section can be
regarded as non-singular supergravity S-branes in the presence of
a positive cosmological constant. As we shall see, solutions in
four and five dimensions can be lifted to supergravity S-brane
configurations in eleven and ten dimensions, respectively.

\section{$D=4$ charged de Sitter bounce}

\subsection{General solution}

For the case of four dimensions, we can add electric and magnetic
charges to the previous de Sitter bounce solution and lift this to
eleven dimensions. We take as our starting point a particular case
of the Petrov type D solution \cite{pleb1,pleb2} which corresponds
to an AdS Reissner-N\"{o}rdstrom Taub-NUT solution. This solution
is given by
\bea ds_4^2 &=&
-\fft{P(r)}{r^2+N^2}\,(dt-2N\,\cos\phi_1\,d\phi_2)^2
+\fft{r^2+N^2}{P(r)}\,dr^2+(r^2+N^2)\,d\Omega_2^2\,,\nn\\
A_\1 &=& \fft{q\,r-N\,p}{r^2+N^2}\,dt-
\fft{p(r^2-N^2)+2N\,q\,r}{r^2+N^2}\,\cos\phi_1\,d\phi_2\,, \eea
\be P(r)=g^2(r^2+N^2)^2+(1+4g^2\,N^2)(r^2-N^2)-2M\,r+q^2+p^2\,,
\ee
and $d\Omega_2^2=d\phi_1^2+\sin^2\phi_1\,d\phi_2^2$. The constant
$N$ is the Taub-NUT parameter, $g$ is the cosmological parameter,
and $q$ and $p$ are the electric and magnetic charges. Taking the
sign of the cosmological constant to be positive (which amounts to
analytically continuing $g\rightarrow {\rm i}\,g$)  and relabeling
$t\rightarrow \psi$ and $r\rightarrow t$ yields
\bea ds_4^2 &=& \fft{{\td
P}(t)}{t^2+N^2}\,(d\psi-2N\,\cos\phi_1\,d\phi_2)^2
-\fft{t^2+N^2}{{\td P}(t)}\,dt^2+(t^2+N^2)\,d\Omega_2^2\,,\nn\\
A_\1 &=& \fft{q\,t-N\,p}{t^2+N^2}\,d\psi-
\fft{p(t^2-N^2)+2N\,q\,t}{t^2+N^2}\,\cos\phi_1\,d\phi_2\,,
\label{charged} \eea
\be {\td
P}(t)=g^2(t^2+N^2)^2+(1-4g^2\,N^2)(N^2-t^2)+2M\,t-(q^2+p^2)\,.
\label{P(t)} \ee

    In order to study the properties of this solution, it is worthwhile
to calculate the Riemann-squared of the metric which is
\bea &&R_{\mu\nu\rho\sigma}\, R^{\mu\nu\rho\sigma} = 24g^4 +
\fft{8}{(t^2 + N^2)^6}\Bigl\{
6 N^2 (4g^2\, N^2 -1) (N^2-t^2) ( N^4 - 14 N^2 t^2 + t^4)\nn\\
&& \ -6M^2(N^2-t^2)(N^4-14N^2 t^2 + t^4)\nn\\
&& \ -24 M\, N^2 (4g^2N^2-1)(N^2-3t^2)(3N^2-t^2)\,t\nn\\ &&
\ +(p^2 + q^2)^2 (7N^4 - 34 N^2\, t^2 + 7t^4)\\
&&\ +12 (p^2 + q^2) [N^2(4g^2N^2 -1)(N^4-10N^2t^2 + 5t^4) -M
(5N^4-10 N^2 t^2 +t^4)t]\Bigr\}\,.\nn \eea
The solution is free of curvature singularities for the entire
coordinate range $-\infty<t<+\infty$.  In the asymptotic regions
$t\rightarrow \pm \infty$, the geometry asymptotes to dS$_4$. In
order for the solution not to have stationary regions with closed
timelike curves, it is necessary for the function ${\td P}(t)$ to
be non-negative everywhere.

     The quartic function $\td P(t)$ has either one or
two minima.  If all the minima are positive, this will ensure that
${\td P}(t)$ is non-negative.  It is convenient to introduce the
function $Q(t)=\td P(t) - (p^2 + q^2)$.  If the lowest minimum of
$Q(t)$, denoted as $Q_{\rm min}$, is non-negative then $\td P(t)$
can be also non-negative, provided that the electric and magnetic
charge parameters $q$ and $p$ satisfy $p^2 + q^2\le Q_{\rm min}$.
Let us assume that the minimum of $Q(t)$ occurs at $t_0$, which
implies
\be
M^2=t_0^2\, (2g^2 t_0^2 + 6 g^2N^2 -1)\,,\qquad
Q_{\rm min} = (t_0^2 + N^2) (1 - 3g^2 N^2 - 3g^2 t_0^2)\,.
\ee
Thus, the condition for $Q_{\rm min}\ge 0$ implies that $g^2 N^2
\le \ft13$ and
\be
M^2 \le M_0^2 \equiv \fft{(1-3g^2N^2) (12g^2 N^2 - 1)^2}{27g^2}\,.
\label{masscons1}
\ee

\begin{figure}
   \epsfxsize=4.0in \centerline{\epsffile{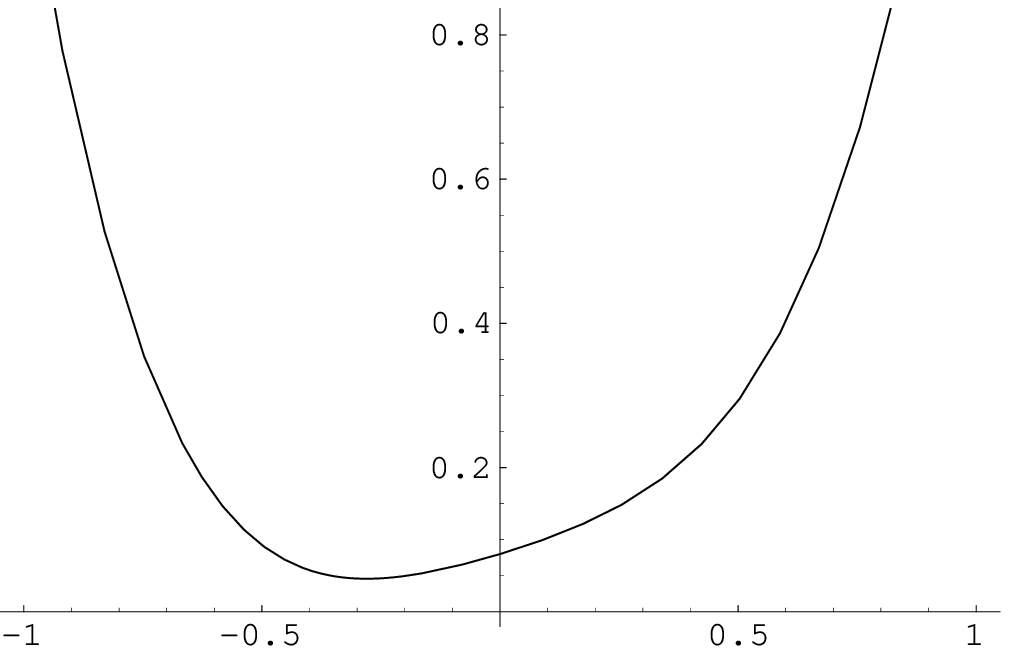}}
   \caption[FIG. \arabic{figure}.]{Function $Q(t)$ for $g=1$, $N^2=\ft15$,
and $M=\ft1{10}$, which has one positive minimum.} \label{bounce4}
\end{figure}

   In the parameter range $\ft16 \le g^2 N^2\le \ft13$, there can only be
one minimum, and hence (\ref{masscons1}) provides the constraint
for the mass parameter $M$.   Figure \ref{bounce4} is the plot of
the function $Q(t)$ for the case with $g=1$, $N^2=1/5$ and
$M=\ft{1}{10}<M_0$ and in this case there is only one minimum.

   When the NUT charge parameter lies in the range
$\ft19 \le g^2 N^2 <\ft16$, there can be one or two minima
depending on the value of $M$.  Introduce a parameter
\be \wtd M_0^2 =\fft{2(1-6g^2 N^2)^3}{27g^2}\,, \ee which is less
than $M_0$ for $g^2 N^2 > \ft19$.  Thus, for $\wtd M_0 \le |M| \le
M_0$, the function $Q(t)$ has only one non-negative minimum.  For
$|M| < \wtd M_0$, $Q(t)$ has two non-negative minima.

\begin{figure}
   \epsfxsize=4.0in \centerline{\epsffile{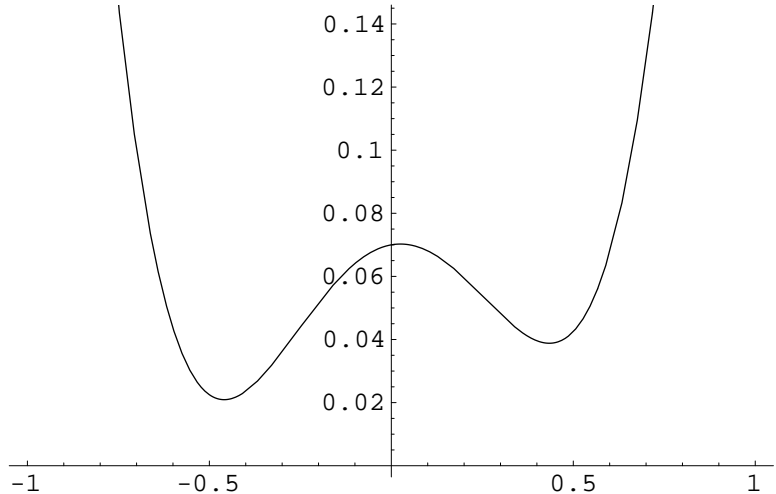}}
   \caption[FIG. \arabic{figure}.]{Function $Q(t)$ for $g=1$, $N^2=\ft1{10}$,
and $M=\ft1{100}$, which has two positive minima.} \label{bounce5}
\end{figure}

     Finally, if $g^2 N^2 <\ft19$, we have
$M_0<\wtd M_0$.  $M$ has to be in the range $|M| < M_0$, for which
there are two positive minima.  As an example, we plot the case
$g=1$, $N^2=\ft1{10}$ and $M=\ft1{100}$ in Figure \ref{bounce5}.

    In summary, for the parameter range discussed above, $\td P(t)$ is
positive-definite and the metric describes a cosmological bounce
between two asymptotic dS$_4$ when $t\rightarrow \pm \infty$. If
we choose the parameters $p$ and $q$ such that the minimum of $\td
P(t)$ is zero at a certain $t_0$, then the metric describes
dS$_2\times S^2$ near the $t_0$ region. The solution thus
interpolates between dS$_2\times S^2$ at the infinite past and
dS$_4$ at the infinite future, with time measured in the co-moving
frame. For the parameters lying outside the above range, the
function $\td P(t)$ can become negative and the solution has
stationary regions with closed timelike curves.

\subsection{BPS solution}

Although the above time-dependent solution is not supersymmetric,
it arises from a first-order system for the cases in which the
corresponding AdS Reissner-N\"{o}rdstrom Taub-NUT solution is
supersymmetric. Thus, the above time-dependent solution might
inherit certain properties of the corresponding supersymmetric
solution, such as stability \cite{cvetic}. The supersymmetry of
the topological AdS Kerr-Newman Taub-NUT solution has been
analyzed in \cite{meessen}. We will focus on the case of the AdS
Reissner-N\"{o}rdstrom Taub-NUT solution, since adding rotation
and then analytically continuing $g\rightarrow {\rm i}\,g$ results
in closed timelike curves. In addition to the Bogomol'nyi bound of
$D=4$ $N=2$ gauged supergravity, given by
\be (M\mp g\,N\,q)^2=(1\pm 2g\,p+5g^2\,N^2)(q^2+p^2)-N^2\, (1\pm
g\,p+4g^2\,N^2)\,, \ee
there is also a supersymmetry constraint relating the charges,
mass and NUT parameter\footnote{This second condition drops out in
the ungauged limit $g=0$.}, given by
\be M\,p=-N\,q\,(1+4g^2\,N^2)\,. \ee

Besides two maximally supersymmetric solutions which are both
locally AdS, there are solutions which preserve a half and a
quarter of the original supersymmetry, for which
\be M_{1/2}=|q \sqrt{1+4g^2\,N^2}|\,,\qquad p_{1/2}=\pm N
\sqrt{1+4g^2\,N^2}\,, \ee
and
\be M_{1/4}=|2g\,N\,q|\,,\qquad p_{1/4}=\fft{1+4g^2\,N^2}{2g}\,,
\ee
respectively \cite{meessen}.

When we analytically continue $g\rightarrow {\rm i}\,g$ to a
time-dependent solution, we find that only the parameter
constraints that originally corresponded to one half preserved
supersymmetry lead to a real time-dependent solution. Again,
although this does not imply that the time-dependent solution is
supersymmetric, it might imply that it is stable. After the
analytical continuation, the physical set of parameter constraints
is
\be M_{1/2}=|q \sqrt{1-4g^2\,N^2}|\,,\qquad p_{1/2}=\pm N
\sqrt{1-4g^2\,N^2}\,, \ee
In order for the time-dependent solution to obey the above
constraints and be free of singularities and closed timelike
curves, we require that
\be 0<g^2\,N^2\le \fft13\,,\qquad |M_{1/2}|\le M_0\,,\qquad
q^2+p_{1/2}^2\le Q_{{\rm min}}\,. \ee
For example, for the single-minimum de Sitter bounce solution
plotted in Figure 4, all of the above conditions are satisfied for
\be g=1\,,\qquad N=\fft{1}{\sqrt{5}}\,,\qquad
M=\fft{1}{10}\,,\qquad q=\fft{1}{2\sqrt{5}}\,,\qquad
p=\fft{1}{5\sqrt{5}} \,. \ee
On the other hand, the two-minimum de Sitter bounce solution
plotted in Figure 5 does not satisfy all of the above constraints.

If the above constraints are to be satisfied, then one can write
\be {\td P}(t)=g^2\,t^4+(6g^2\,N^2-1)t^2+2q\sqrt{1-4g^2\,N^2}t+
g^2\,N^4-q^2\,.
\ee
In order to have a real solution, we must have $g\,N\le 1/2$. The
analysis for the requirement that $\td P(t)$ be non-negative is
analogous to that of the more general non-BPS solution that we
discussed earlier. Whether the function $\td P(t)$ has one or two
minima depends on the parameters $q$ and $N$. Suppose that the
minimum $P_{\rm min}$ occurs at $t_0$. Then we have
\be q^2=\fft{(8g^2 N^2-1)^2}{16g^2}\,,\qquad \td P_{\rm
min}=\fft{g^2(N^2+t^2)^2(1 - 4g^2 N^2 - 4g^2 t_0^2)}{ 1-4g^2
N^2}\,. \ee
Thus, it follows that for $\td P_{\rm min} \ge 0$, we have $g^2
N^2\le \ft14$ and
\be q^2 \le q_0^2 \equiv \fft{(8g^2 N^2-1)^2}{16g^2}\,. \ee

     In the parameter range $\ft16 \le g^2 N^2 \le \ft14$, there can
only be one minimum.  When the NUT charge parameter lies in the
range $\ft{5}{36} \le g^2\, N^2 < \ft16$, there can be either one
or two minima, depending on the value of $q$.  We define $\td
q_0^2 = 2 (1-6g^2N^2)^3/(27 g^2 (1 - 4g^2 N^2))$, which is smaller
than $q_0^2$ in the above range. The function $\td P(t)$ has a
single non-negative minimum if $\td q_0\le |q|\le q_0$ and two
positive minima instead if $|q|<\td q_0$.  For $g^2N^2 <\ft5{36}$,
we have $q_0 < \td q_0$ and $\td P(t)$ remains non-negative,
provided that $|q|\le q_0$.

     A second-order zero arises when $P_{\rm min}=0$, which can occur
when $|q|=q_0$. In this case, the solution interpolates between
dS$_2\times S^2$ in the infinite past to dS$_4$ in the infinite
future, where time is measured in the comoving frame.

\section{De Sitter bounces in M-theory}

\subsection{Lifting the four-dimensional bounce}

We can lift the four-dimensional charged de Sitter bounce to
eleven dimensions on a seven-dimensional hyperbolic space
\cite{lvpds}. The resulting $SO(4,4)$ gauged solution is given by
\bea ds_{11}^2 &=& \Delta^{2/3}\, \Big( \fft{{\td
P}(t)}{t^2+N^2}\,(d\psi-2N\,\cos\phi_1\,d\phi_2)^2
-\fft{t^2+N^2}{{\td P}(t)}\,dt^2+(t^2+N^2)\,d\Omega_2^2\Big) \nn\\
&& +\fft{1}{\Lambda}\, \Delta^{2/3}\,d\theta^2+
\fft{1}{4\Lambda}\,\Delta^{-1/3}\,\Big[c^2\,\Big( d{\overline
\Omega}_2^2+({\overline
\sigma}_3-\sqrt{\Lambda}A_\1)^2\Big)\nn\\
&& +s^2\,\Big( {\td\Omega}_2^2+
({\td\sigma_3}-\sqrt{\Lambda}A_\1)^2\Big)\Big]\,,\nn\\
F_\4 &=&
2\sqrt{\Lambda}(t^2+N^2)\,(d\psi-2N\,\cos\phi_1\,d\phi_2)\wedge
dt\wedge \Omega_\2\nn\\ && -\fft{1}{4\Lambda}\Big(
sc\,d\theta\wedge ({\overline \sigma}_3-{\td \sigma}_3)-\fft12
c^2\,{\overline \Omega}_\2+\fft12 s^2\,{\td \Omega}_\2\Big)\wedge
\ast F_\2 \,, \label{44} \eea
where $\Delta =\cosh 2\theta$, $c=\cosh\theta$ and
$s=\sinh\theta$. $\sigma_i$ and $\td \sigma_i$ are $SU(2)$
left-invariant 1-forms satisfying
\be d\sigma_i =-\ft12
\epsilon_{ijk}\,\sigma_j\wedge\sigma_k\,,\qquad d\td\sigma_i =
-\ft12 \epsilon_{ijk}\, \td\sigma_j\wedge\td\sigma_k\,. \ee
$F_\2=dA_\1$, where $A_\1$ is given in (\ref{charged}). ${\td
P}(t)$ is given in (\ref{P(t)}).

The squashed $S^3$ of the four-dimensional bounce and one of the
internal $S^3$ have strictly positive factors. This ensures that,
upon replacing each of these $S^3$ by a three-dimensional lens
space $S^3/Z_Q$ and reducing and T-dualizing over the two fibre
coordinates, the resulting time-dependent solution of type IIB
theory is non-singular\footnote{This procedure of replacing
3-spheres by lens spaces has been used to find new warped
embeddings of AdS \cite{warped}.}.

For vanishing electric and magnetic charges, we can also embed
the solution in eleven dimensions as an $SO(5,3)$ gauged solution
\cite{hullgibbons}, given by
\bea ds_{11}^2 &=& \td\Delta^{2/3}\, (-\fft{1+t^2}{\td
P(t)}\,dt^2+\fft{\td P(t)}{1+t^2}\, (d\psi-2\td
A)^2+(1+t^2)\,d\Omega_2^2)\nn\\ && +\fft{1}{\Lambda}\,
\td\Delta^{-1/3}\,\Bigl( \fft19 (2\cosh(2\theta)+1)\, d\theta^2+
\cosh^2 \theta\,d\Omega_4^2+\fft13 \sinh^2
\theta\,\td\Omega_2^2\Bigr) \,,\nn\\
F_\4 &=& 2\sqrt{\Lambda}\,\epsilon_\4\,. \label{53} \eea
\be \td\Delta=\fft13 (5\cosh (2\theta)-2)\,. \ee

\subsection{Lifting the five-dimensional bounce}

Five-dimensional solutions are given by (\ref{bounce1n}) and
(\ref{P1}), for which $m=1$ and $n=1$. These include solutions
which smoothly run from dS$_3\times \Sigma_2$ to dS$_5$, as well
as bounces between two phases of dS$_5$. These solutions can be
embedded in ten-dimensional type IIB theory as
\bea ds_{10}^2 &=& \Delta^{1/2}\, \Big( -\fft{1+t^2}{\td
P(t)}\,dt^2+\fft{\td P(t)}{1+t^2}\,
(d\psi-2A)^2+(1+t^2)\,d\Omega_2^2+t^2\, d\phi^2\Big) \nn\\
&& + \fft{1}{\Lambda}\, \Delta^{1/2}\,d\theta^2 +
\fft{1}{\Lambda}\,\Delta^{-1/2}\,(\cosh^2
\theta\,d\Omega_2^2+\sinh^2
\theta\,\td\Omega_2^2)\,,\nn\\
F_\5 &=& 2\sqrt{\Lambda}\,(\epsilon_\5+\ast \epsilon_\5)\,,
\label{33} \eea
where $\Delta =\cosh 2\theta$ and $\epsilon_\5$ is the volume-form
corresponding to the five-dimensional metric. We have set $c$ and
$P_0$ to unity for simplicity.

As before, we can replace the $S^3$ of the five-dimensional metric
by $S^3/Z_q$. The solution can then be Hopf T-dualized and lifted
to yield a non-singular time-dependent solution in eleven
dimensions.

\section{Conclusions}

The analytical continuation of recently-found de Sitter instantons
leads to asymptotically de Sitter cosmological solutions, which
include smooth de Sitter bounces. In four dimensions, we were able
to generalize this construction to obtain a de Sitter bounce with
electric and magnetic charges, by analytically continuing the AdS
Reissner-N\"{o}rdstrom Taub-NUT solution. Four and
five-dimensional de Sitter bounces can be lifted to non-singular
time-dependent configurations in eleven-dimensional supergravity
and type IIB theory, respectively. These bounces are completely
regular in a purely classical gravity setting. They asymptotically
approach de Sitter spacetime in both the infinite past and
infinite future but deviate from de Sitter during intermediate
times, with respect to the comoving time.  It is clear that
further analytical continuations of these solutions can be
performed, which may lead to more non-singular solutions. This
will be discussed elsewhere.

These solutions are potentially interesting new backgrounds on
which to study dS/CFT-type holography \cite{strominger1} in the
case where the conformal symmetry is broken throughout the
cosmological evolution and restored at very early and late times.
It has been postulated that a cosmological flow corresponds to a
renormalization group flow between two conformal fixed points of a
three-dimensional Euclidean field theory \cite{strominger2}.
However, our cosmological solution which runs from dS$_2\times
S^2$ to dS$_4$ corresponds to a flow ``across dimensions'' from a
one-dimensional Euclidean field theory to a three-dimensional
Euclidean field theory\footnote{The AdS/CFT analog of such flows
has been discussed in, for example, \cite{nunez,cucu1,cucu2}.}.
The solution can be fine-tuned to expand quickly from the
dS$_2\times S^2$ fixed-point. In this case, the early-time
behavior might not be completely unrealistic. In fact, this
fixed-point might leave an observable imprint on the CMB as a
contribution to the quadrupole moment\footnote{We would like to
thank Stephon Alexander for this suggestion.}. The solution can be
fine-tuned to describe an expanding universe whose expansion rate
is significantly larger in the past than in the future, providing
an inflationary model with no singularity. With regards to the de
Sitter bounces, it has been postulated that the Euclidean field
theory on only one of the boundaries is physically relevant
\cite{strominger1}.

It is interesting to study the dampening effects of the
cosmological constant in our solutions. For vanishing mass, NUT
charge, and electric and magnetic charges, the cosmological bounce
reduces to pure de Sitter in global coordinates.  As the other
parameters are turned on, the effect of the cosmological constant
can be reduced. If this dampening effect occurred at late times
then this could have some interesting ramifications with regards
to the cosmological constant problem. The mass parameter breaks
the symmetry of the bounce solution about the central region.
Thus, we postulate that the cosmological bounce could be
fine-tuned to have a longer period of slow contraction. If this is
the case, then the cosmological bounce is a viable alternative to
inflation. In particular, during the end of the contracting phase,
currently-observed cosmological scales would be well inside the
Hubble radius, providing ample time for the observed homogeneity
to occur through causal microphysics.

\section*{Acknowledgments}

We would like to thank Vijay Balasubramanian, Mirjam Cveti\v{c}
and Proty Wu for discussions. H.L. is supported in part by DOE
grant DE-FG03-95ER40917. J.F.V.P. is supported in part by DOE
grant DE-FG01-00ER45832. J.E.W. is supported in part by the
National Science Council, the Center for Theoretical Physics at
National Taiwan University, the National Center for Theoretical
Sciences, and the CosPA project of the Ministry of Education.

\end{document}